\documentclass[
    twocolumn,
	prd,
	amssymb,
	preprintnumbers,superscriptaddress,
	nofootinbib]{revtex4-1}

\pdfoutput=1
\usepackage{paralist}
\usepackage{graphicx}
\usepackage{enumitem}
\usepackage{latexsym}
\usepackage{amsfonts}
\usepackage{amssymb}
\usepackage{xcolor}
\usepackage[export]{adjustbox}
\usepackage{amsmath}
\usepackage{slashed}
\usepackage{dcolumn}
\usepackage{verbatim}
\usepackage{float}
\usepackage{multirow}
\usepackage{xspace}
\usepackage[normalem]{ulem}
\usepackage[
pdfauthor={Jeremy Sakstein}]{hyperref}
\usepackage{tabularx}

\newcommand{\beq}{\begin{equation}}
\newcommand{\eeq}{\end{equation}}
\newcommand{\bea}{\begin{eqnarray}}
\newcommand{\eea}{\end{eqnarray}}

\newcommand{\pL}{\left(} \newcommand{\pR}{\right)} \newcommand{\bL}{\left[} \newcommand{\bR}{\right]}

\newcommand{\refjnl}[1]{{\rm#1}}

\def\apjl{\refjnl{ApJ}}                
\def\aap{\refjnl{A\&A}}                

\def\prd{\refjnl{Phys.~Rev.~D}}        
\newcommand{\msun}{{\rm M}_\odot}

 \newcommand{\bP}{{ \mathbb P}}

\newcommand{\mbh}{M_{\rm BH}}
\newcommand{\mbhmg}{M_{\rm BHMG}}
\newcommand{\mmin}{M_{\rm min}}

\newcommand{\FAR}{\text{FAR}}

\DeclareRobustCommand{\okina}{%
  \raisebox{\dimexpr\fontcharht\font`A-\height}{%
    \scalebox{0.8}{`}%
  }%
}

\allowdisplaybreaks

\begin{document}

\title{Multi-Generational Black Hole Population Analysis with an Astrophysically Informed Mass Function}
\author{Yannick Ulrich} \email{yannick.ulrich@cern.ch}
\affiliation{Institute for Particle Physics Phenomenology, Department of Physics, Durham University, Durham DH1 3LE, U.K.}
\affiliation{Institut für Theoretische Physik \& AEC, Universit\"at Bern, Sidlerstrasse 5, CH-3012 Bern, Switzerland}
\author{Djuna Croon} \email{djuna.l.croon@durham.ac.uk}
\affiliation{Institute for Particle Physics Phenomenology, Department of Physics, Durham University, Durham DH1 3LE, U.K.}
\author{Jeremy Sakstein} \email{sakstein@hawaii.edu}
\affiliation{Department of Physics \& Astronomy, University of Hawai\okina i, Watanabe Hall, 2505 Correa Road, Honolulu, HI, 96822, USA}
\author{Samuel D. McDermott}
\email{samueldmcdermott@gmail.com}
\affiliation{Department of Astronomy and Astrophysics,
University of Chicago, Chicago, IL, USA}

\date{\today}

\begin{abstract}
We analyze the population statistics of black holes in the LIGO/Virgo/KAGRA GWTC-3 catalog using a parametric mass function derived from simulations of massive stars experiencing pulsational pair-instability supernovae (PPISN).~Our formalism enables us to separate the black hole mass function into sub-populations corresponding to mergers between objects formed via different astrophysical pathways, allowing us to infer the properties of black holes formed from stellar collapse and black holes formed via prior mergers separately.~Applying this formalism, we find that this model fits the data better than the powerlaw+peak model with Bayes factor $ 14.3\pm0.1$.~We measure the location of the lower edge of the upper black hole mass gap to be $\mbhmg=70.09_{-14.72}^{+19.82}\msun$.~
We combine our mass function with the spectral siren method for measuring the Hubble constant to find $H_0={94.62^{+30.24}_{-30.83}}$ km/s/Mpc.~Our results demonstrate how astrophysically-informed mass functions can facilitate the interpretation of gravitational wave catalog data to provide  information about black hole formation and cosmology.
\end{abstract}

\preprint{IPPP/24/29}

\maketitle

\section{Introduction}

The LIGO/Virgo/KAGRA (LVK) gravitational wave interferometers have detected a rapidly growing catalogue of events.
~Interpreting the data in the context of the underlying physics requires knowledge of the  black hole mass function (BHMF), which is presently unknown due to the difficulty in predicting it from first principles.~The majority of objects observed by LVK are expected to be \textit{first-generation} black holes (BHs) formed from the collapse of massive stars ($M\gtrsim20\msun$), but  some fraction may be \textit{second generation} black holes (or higher-generation) that formed from the merger of two first generation objects.
~By fitting parameterized BHMFs to the data, the physics underlying the formation of these diverse populations can be inferred.~The free parameters in the BHMFs provide measurable constraints on the formation processes.

Another application of BHMFs is to measure cosmological parameters via the \textit{spectral sirens} method.~LVK measures the detector frame masses of each black hole and the luminosity distance to the merger, but the redshift is unknown.~Features in the BHMF break the degeneracy between redshift and source-frame mass, enabling measurements of various cosmological parameters.~Of these, the Hubble constant $H_0$ is best-constrained using this method,
but the measured value is highly-sensitive to the assumed BHMF \cite{Pierra:2024wrk}.  

In reference \cite{Baxter:2021swn}, three of us proposed a novel BHMF which models the physics of pair-instability supernovae with three free parameters.~This mass function was designed to capture the physics of pulsational pair instability supernovae (PPISN), which predicts an \textit{upper black hole mass gap} (BHMG) --- a discontinuity in the black hole mass distribution preceded by a peak due to PPISN pile-up \citep{2016A&A...594A..97B, 2017ApJ...840L..24F, 2017PhRvD..95l4046G, Talbot:2018cva, Wang2020arXiv200903854W, Wang2021arXiv210402566W}.~Bayesian analysis of the GWTC-2 catalog with this mass function resulted in mass gap $\mbhmg=55.4^{+3.0}_{-6.1}\msun$, compatible with the expectation from  stellar structure theory.~The higher-generation BHs that \textit{pollute} the gap were modeled as a single additional background; luminosity distance information was not used, and, aligned with conventions in the field, only the heaviest black hole in each merger event (the ``primary'') was fit using the mass function.

In this work, we extend the formalism and analysis in \cite{Baxter:2021swn} in two key directions that connect population modeling more directly to both formation physics and cosmology. 
First, we move beyond primary-only analyses and model the joint distribution of primary and secondary masses, introducing a multi-component BHMF that explicitly accounts for higher generation BHs.
This allows us to constrain the fraction of mergers involving hierarchical assembly and to test whether features such as the PISN gap are consistently present across both components.
Secondly, we incorporate luminosity distance information to measure the Hubble constant using the spectral sirens method.

Although this analysis is based on GWTC-3 \cite{LIGOScientific:2021djp}, our goal is to develop and test a physically motivated framework for interpreting black hole populations.~This approach is complementary to analyses of more recent catalogs.~Several studies of the newly released GWTC-4 have revisited evidence for features associated with pair-instability physics in the black hole mass distribution. While these analyses generally report evidence for an upper mass gap and possible excesses consistent with stellar evolution theory \cite{Tong:2025wpz,Guttman:2025jkv,Banagiri:2025dmy}, the inferred significance and interpretation of these features depend sensitively on the assumed population model and treatment of hierarchical mergers \cite{Afroz:2025xpp,Sridhar:2025kvi,Tiwari:2025oah,Willcox:2025cus}. 
In this work, we instead model the joint distribution of $(M_1,M_2)$ directly using a multi-component mass function that includes contributions from both first- and higher-generation BH. This framework allows both components to be drawn from mixtures of formation channels, with correlations encoded through the mass-ratio distribution, and provides a complementary approach for assessing the robustness of inferred features in the mass spectrum.

A summary of our key results is as follows:
\begin{compactitem}
    \item Our PPISN-informed BHMF is preferred over the phenomenological power law+peak model with Bayes factor $\log Z_{\rm PPISN} - \log Z_{\rm PLP} =14.3\pm0.1$.
    \item We measure the location of the upper BHMG to be $\mbhmg=70.09_{-14.72}^{+19.82}\msun$, larger than previous measurements but consistent with stellar structure theory. 
    \item We measure $H_0=94.62^{+30.24}_{-30.83}$ km/s/Mpc, consistent with astrophysical and cosmological measurements at the $\sim$$2\sigma$ level.
\end{compactitem}
We discuss these findings and their origin in detail below.

This paper is organized as follows.~In Section~\ref{sec:massfn} we introduce the mass function we use in our data analysis, and describe the spectral sirens method for measuring $H_0$.~In Section~\ref{sec:data} we describe our data analysis.~Our results are presented in Section~\ref{sec:results}.~We discuss these results and conclude in Section~\ref{sec:discussion}.

\section{Population Modeling}\label{sec:massfn}

\subsection{Mass function}
\label{sec:massFunction}

In what follows we assume the primary and secondary black holes are picked from identical mass distributions, and that the different generations of black holes are represented by different mass distributions.~This implies, using the chain rule of probability, that the total probability distribution is given by 
\begin{equation}
    \begin{split}
        \bP(\vec \omega| \vec \theta) \propto& p^{(1{\rm g})}(M_1 | \vec \theta) p^{(1{\rm g})}(M_2 | \vec \theta) f(q;\beta^{(1)}_q) \\  &+ \lambda_{12}p^{(1{\rm g})}(M_1 | \vec \theta) p^{(2{\rm g})}(M_2 | \vec \theta) f(q;\beta^{(2)}_q)
\\  &+ \lambda_{21}p^{(2{\rm g})}(M_1 | \vec \theta) p^{(1{\rm g})}(M_2 | \vec \theta)  f(q;\beta^{(2)}_q) + \cdots .~
    \end{split}
    \label{eq:multipop}
\end{equation}
where 
$p^{(n{\rm g})} = \frac{dN^{(n{\rm g})}}{dM}$ is the underlying mass function of the $n^{\rm th}$ generation black holes.~Thus, the different terms account for BH binaries formed from a stellar binary, and BH binaries which include a higher generation BH. We assume that the primary formation channel for BBH mergers is from stellar collapse, implying that the first-generation black holes dominate the mass function. In \eqref{eq:multipop}, the function $f$ accounts for a preferential formation mechanism for binaries with a particular mass ratio $q\equiv M_2/M_1$.~Following \cite{Farah:2023swu}, we choose $ f = q^{\beta_q}$.~We allow for different $\beta_q$ between binaries formed of two first generation black holes.

The mass function we use for the first generation black holes is \cite{Baxter:2021swn}
\begin{equation}\label{nbh-1g}
\begin{split}
    p^{(1{\rm g})}
    \propto\,\,& \! \mbh^b \! \bL 1 \! + \! \frac{2 a^2 \mbh^{1/2} 
(\mbhmg - \mbh)^{a-1}}{ \mbhmg^{a-1/2}} \bR\! \\ 
& \times  S(\mbh|\mmin,\delta_m)
\end{split}
\end{equation}
with ``turn-on" function at the low mass end, 
\begin{equation} \label{eq:smooth-turn-on}
\begin{split}
        S(x|y,z)
    &=  \bL\exp \pL \frac{z}{x-y} + \frac{z}{x-y-z} \pR +1 \bR^{-1},
    \\
    & \int_y^{y+z}dx S(x|y,z) = \frac z2
\end{split}
\end{equation}
for black holes in the range $\mmin \leq \mbh \leq \mmin+\delta_m$, and $S=0 (1)$ for $\mbh<\mmin$ ($\mbh>\mmin + \delta_m$), following the analysis in \cite{LIGOScientific:2020kqk}.~This mass function was derived from one-dimensional stellar structure simulations of stellar collapse to black holes and incorporates the effects of PPISN.~Reference \cite{Baxter:2021swn} found that this was a good fit to the BHMF for a wide range of uncertain stellar parameters e.g., nuclear reaction rates.~The parameter $\mbhmg$ corresponds to the location of the BHMG, the parameter $a$ measures the width of the black hole mass spectrum (final BH mass vs.~initial BH mass), and the parameter $b$ measures a combination of the slope of the initial mass function (IMF) and the zero-age main-sequence to zero-age helium burning transfer function.~Note that this function diverges at $M=\mbhmg$, forcing the peak in the black hole mass spectrum (the heaviest BH formed from stellar collapse) to correspond to the location of the upper BHMG. See \cite{Baxter:2021swn} for more details.

We assume that the higher generation black holes are increasingly rarer and that binary formation mechanisms for black holes from the same generation are not more efficient than cross-generational coupling.~We therefore enforce a hierarchy in $\lambda_{ij}$ parameters where 
$1 > \lambda_{21} > \lambda_{12} > \cdots$.~We will ignore third-order and higher generations in the following 
so we only need to include a distribution for the second-generation black holes.~As in~ \cite{Baxter:2021swn}, we use
\begin{equation}
\label{eq:2gMF}
\begin{split}
        p^{(2{\rm g})} =& S(\mbh|\min,\delta_m) \\
        &
        \begin{cases}
            1 & m < m_{\rm 2g \, min} \\
            \Big(\frac{m}{\mbhmg + \mmin+\delta_m/2}\Big)^d & m \geq  m_{\rm 2g \, min}
        \end{cases}
\end{split}
\end{equation}
where $m_{\rm 2g \, min} = \mbhmg + \mmin+\frac{\delta_m}{2}$.~This approximates the secondary mass function found in Figure~5 of \cite{Baxter:2021swn} in the absence of mass- and environment-dependent binary formation efficiency effects.~We will explicitly verify that our assumption $\lambda_{ij} \ll 1$ is satisfied \textit{a posteriori}.

The $\lambda$s in \eqref{eq:multipop} are a measure of the number black holes in each population, provided that the different contributions to \eqref{eq:multipop} are appropriately normalized.
To be precise, we write
\begin{equation}
    \lambda_{12} = \frac{N(2\in 2{\rm g})}{N_{\rm tot}}
    \qquad\text{and}\qquad
    \lambda_{21} = \frac{N(1\in 2{\rm g})}{N_{\rm tot}}\,,
\end{equation}
where $N(i\in 2{\rm g})\equiv N_i$ counts the number of primary ($i=1$) and secondary ($i=2$) black holes in the second generation.~To avoid the dependence on $N_{\rm tot}$ we similarly introduce $N_0$  that counts the number of events with both black holes being first generation.~With this, we now need the calculate the following normalization integrals
\begingroup\makeatletter\def\f@size{9}\check@mathfonts
\def\maketag@@@#1{\hbox{\m@th\large\normalfont#1}}%
\begin{align}\begin{split}
    \frac{N_{\rm tot}}{N_0} = \int_0^\infty dm_1\ \int_0^{m_1}dm_2\ p^{(1{\rm g})}(M_1 | \vec \theta) p^{(1{\rm g})}(M_2 | \vec \theta) f(q;\beta^{(1)}_q) \\
    \frac{N_{\rm tot}}{N_1} = \int_0^\infty dm_1\ \int_0^{m_1}dm_2\ p^{(1{\rm g})}(M_1 | \vec \theta) p^{(2{\rm g})}(M_2 | \vec \theta) f(q;\beta^{(2)}_q)\\
    \frac{N_{\rm tot}}{N_2} = \int_0^\infty dm_1\ \int_0^{m_1}dm_2\ p^{(2{\rm g})}(M_1 | \vec \theta) p^{(1{\rm g})}(M_2 | \vec \theta)  f(q;\beta^{(2)}_q)
    \end{split}
\end{align}
These integrals can be split into terms that can be calculated analytically in terms of hypergeometric functions and those that involve the smoothing function $S$ which cannot be analytically integrated.
The five remaining integrals (three one dimensional and two dimensional ones) are evaluated numerically over a fine grid as part of the inference process. We have provided a machine-readable version of these integrals as an ancillary file to this submission\footnote{\href{https://doi.org/10.5281/zenodo.15608528}{https://doi.org/10.5281/zenodo.15608528}
} \cite{zenodo}.

We will compare our mass function to the phenomenological power law+peak (PLP) mass function introduced in \cite{LIGOScientific:2020kqk,LIGOScientific:2018jsj}.~This mass function was also motivated by the expectation of a PPISN pile-up, and provides the best fit to the GWTC-2 and GWTC-3 of the models studied by LVK \cite{Abbott:2020gyp,LIGOScientific:2021psn}.~The primary mass distribution follows, 
\begin{equation}
\label{eq:plp}
\begin{split}
     p (M_1  | \vec \theta ) =&
    \bigg[
    (1-\lambda_\text{peak})\mathfrak{P}(m_1|-\alpha, m_{\rm max})  
     \\ 
    &+ 
    \lambda_\text{peak} G(m_1|\mu_m,\sigma_m)
    \bigg] S(m_1|\mmin, \delta_m) .~  
\end{split}
\end{equation}
where $\mathfrak{P}(m_1|-\alpha,m_{\rm max})$ is a normalized power law distribution with spectral index $-\alpha$ and high-mass cut-off $m_{\rm max}$, $G(m_1|\mu_m, \sigma_m)$ is a normalized Gaussian distribution with mean $\mu_m$ and width $\sigma_m$, the parameter $\lambda_\text{peak}$ is a mixing fraction determining the relative prevalence of mergers in $\mathfrak{P}$ and $G$, and  $S(m_1, \mmin, \delta_m)$ is the ``turn-on'' function \eqref{eq:smooth-turn-on}. 

In order to capture possible cosmic evolution of the binary merger rate, we model the source‐frame merger‐rate density as
\begin{equation}
  R(z)\,\propto\,(1+z)^\gamma,  
\end{equation}
so that, when mapping from detector‐frame distances to source‐frame rates, the usual geometric factor $d_L^2/H(z)$ is modified by an additional weight $(1+z)^{\gamma-3}$.
Thus $\gamma=0$ corresponds to a merger-rate density that is constant in comoving coordinates, while $\gamma>0$ ($\gamma<0$) describes a rate that increases (decreases) with redshift. For $\gamma=3$, the intrinsic rate evolution exactly cancels the $(1+z)^{-3}$ factor arising from the comoving-volume conversion and cosmological time dilation.

\subsection{Spectral Sirens}
\label{sec:spectralSirens}
The spectral siren method can be used to make a measurement of the Hubble constant $H_0$ using features in the BHMF without localizing each event's host galaxy~\cite{Ezquiaga:2021ayr,Ezquiaga:2022zkx}.~The method works by assuming an underlying BHMF.~This breaks the degeneracy between BH mass and source redshift, allowing for a measurement of cosmological parameters when the latter is unknown.~Unfortunately, assumptions about the source population strongly impact the determination of $H_0$ \cite{Mastrogiovanni:2021wsd,Pierra:2023deu}.~As such, phenomenological mass functions introduce a degree of ambiguity in the determination of $H_0$, which astrophysically-motivated mass functions such as ours may help to alleviate.

To apply this method we use the detector frame masses and the luminosity distance, and solve
\begin{equation}
\label{eq:DL}
        D_L(z) = \frac{(1+z)}{H_0} \int_0^z \frac{dz'}{ \sqrt{\Omega_m (1 + z')^3 + (1 - \Omega_m)}}
\end{equation}
to find $z$ for each event.~Note that we have assumed flat $\Lambda$CDM in Eq.~\eqref{eq:DL}.~We then calculate the source frame masses from the detector frame masses as
\begin{equation}
    m_{1,2_{\text{src}}} = \frac{m_{1,2_{\text{det}}}}{1+z}.
\end{equation}
In principle, there are two free parameters, $H_0$ and $\Omega_m$, but in practice $\Omega_m$ has a negligible impact on the estimate of $H_0$ \cite{Mastrogiovanni:2021wsd} so we fix $\Omega_m=0.308$, the best-fit value found by the Planck collaboration \cite{Planck:2015fie}.

\begin{table}[t]
\centering
\begin{tabular}{|c|c|}
\hline
\multicolumn{2}{|c|}{Mass Function (Eqs.~\eqref{eq:multipop}, \eqref{nbh-1g}, and \eqref{eq:2gMF})}    \\\hline
$\mbhmg$   & $[20\msun, 120\msun]$ \\\hline
$a$        & $[0,+1/2]$ \\\hline
$b$        & $[-4,0]$ \\\hline
$\log_{10}(\lambda_{12})$,   $\log_{10}(\lambda_{21})$ & $[-7,-0.3]$\\\hline
$d$ & $[-10,0]$\\\hline
$M_{\rm min}$ & $[2\msun,10\msun]$\\\hline
$\delta_m$ & $[0\msun,10\msun]$\\\hline
{$\beta_q$} & $[-4,12]$\\\hline
$\gamma$ & $[-6, 6]$\\\hline
\multicolumn{2}{|c|}{Spectral Sirens}
\\\hline  
$H_0$ (km/s/Mpc) & $[{20,150}]$\\\hline
\end{tabular}%
\caption{Prior ranges for the parameters used in our data analysis.} 
\label{tab:priors}
\end{table}

\section{Data Analysis}
\label{sec:data}

We have implemented the models described in Section~\ref{sec:massfn} into a new code:~MULTIPASS --- MULTI-Population Analysis of Stellar and Secondary black holes --- which uses the FORTRAN library Multinest~\cite{Feroz:2008xx} to perform the Bayesian inference.~Our code is available at~\cite{zenodo} and on GitHub\footnote{\url{https://github.com/multipass-black-holes/multipass}.}
The code's modular design makes it possible to easily add new mass functions or features.~Together with the code we distribute our data preparation scripts (see below) and run cards for  reproducibility.~We have verified that the code gives  results compatible with the LVK analysis \cite{LIGOScientific:2021psn}. 

We use the injection campaign of~\cite{injection} with a false-alarm rate (FAR) filter of $\FAR < 1$, reweighting the injections using the provided sampling probability distribution functions.~Similarly, we select all events up to and including GWTC-3 with $\FAR < 1$ except GW170817 (a binary neutron star (BNS) merger), GW190814 and GW190917\_114630 (which include a lower mass gap object), GW190426\_152155 (large false alarm rate), GW190425 (likely BNS), and GW200115\_042309 (NSBH merger).~This selection leaves us with 69 merger events, a detailed list of which can be found in~\cite{zenodo}.

The parameters and prior ranges we use are given in table \ref{tab:priors}.~In the case of the mass function, the prior ranges for the  parameters in Eq.~\eqref{nbh-1g} are identical to those used in \cite{Baxter:2021swn}, with the parameters in the smoothing function being those chosen by LVK \cite{LIGOScientific:2020kqk}.
~In the case of the spectral sirens analysis, we do not normalize the different mass functions $p^{(n{\rm g})}$ because the measurement on $H_0$ is only sensitive to features in the mass function and does not depend on the generation of each black hole.~We  check \textit{a posteriori} that the hierarchy in $\lambda_{ij}$ motivated above is respected such that it remains appropriate to truncate the mass function.

\begin{figure}[t]
    \centering
    \includegraphics[width=.5\textwidth]{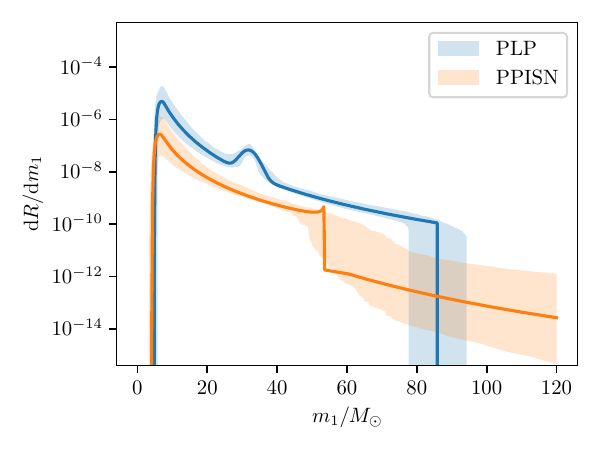}
    \caption{95$\%$ confidence intervals of the mass function for the primary black hole in the  power law+peak model \eqref{eq:plp}, and the mass function inspired by simulations of PPISN and multiple populations \eqref{eq:multipop} presented here. 
    }
    \label{fig:masscontours}
\end{figure}

\section{Results}
\label{sec:results}

\subsection{Mass Function Analysis}
\label{sec:massDistribution}

\begin{figure*}
    \centering
    \includegraphics[width=\textwidth]{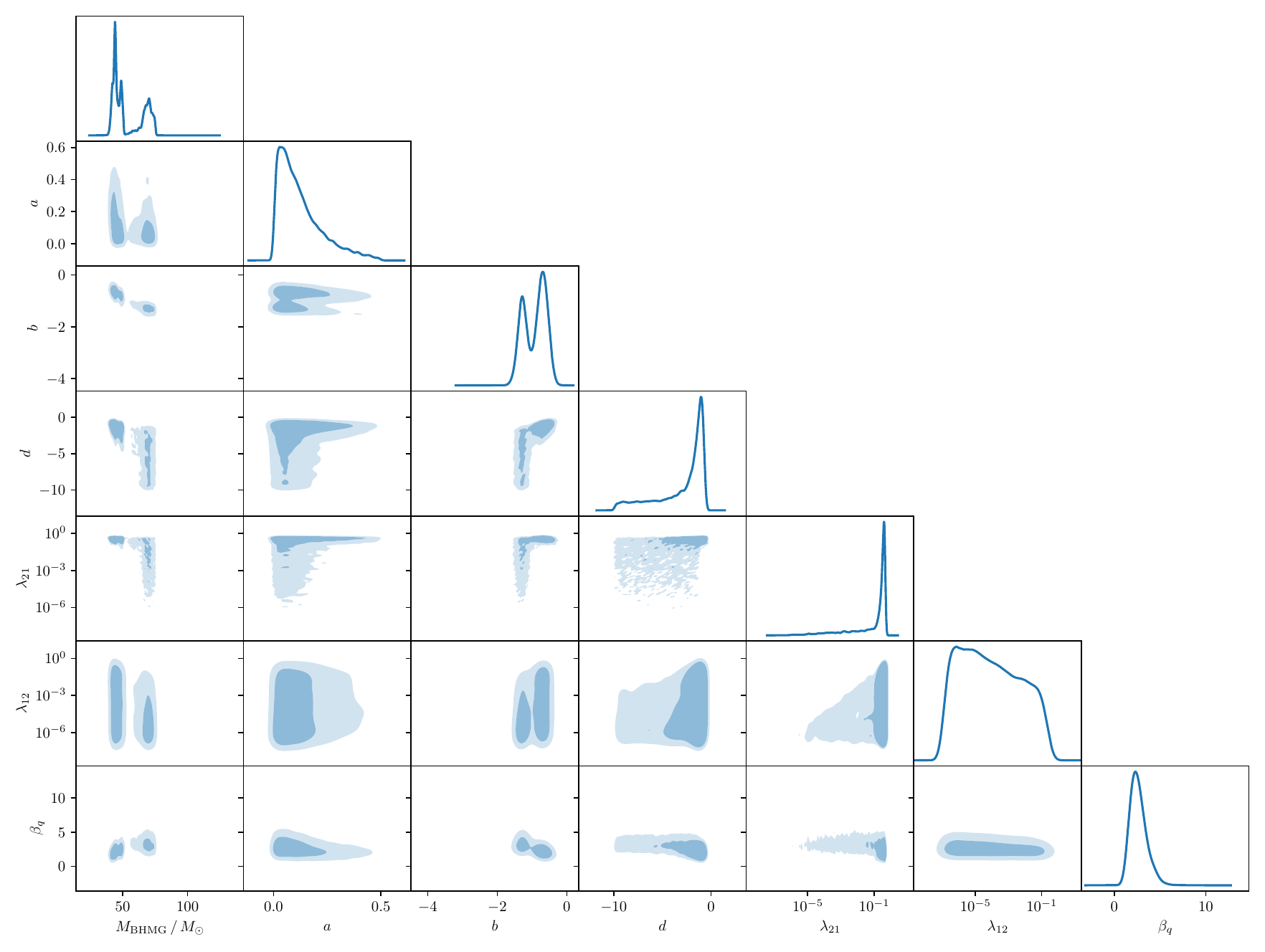}
    \caption{ 
    Corner plot of the parameters in the mass function in equations~\eqref{eq:multipop}, \eqref{nbh-1g}, and \eqref{eq:2gMF}.
    ~The variation in each parameter depicted corresponds to the prior range used in our study.
    ~Parameters of the mass function ``turn-on'' are not depicted, but do not correlate significantly with any of the parameters shown.
    ~It is seen that the distribution in the mass gap parameter $\mbhmg$ is bimodal, with the most important support around $\mbhmg\sim 50 \msun$, consistent with stellar structure theory.
    }
    \label{fig:spin-corner}
\end{figure*}

The results of fitting our mass function in equation~\ref{nbh-1g} to the GWTC-3 catalog (sans the events stated above) using our split-population formalism in equation~\eqref{eq:multipop} are shown in Figure~\ref{fig:masscontours} and in the corner plots in  Figure~\ref{fig:spin-corner};~the best-fitting parameters are given in table~\ref{tab:bestfit-massfct}.~Our mass function defined by \eqref{eq:multipop}, \eqref{nbh-1g} and \eqref{eq:2gMF} provides a better fit to the data than PLP, with a Bayes evidence factor of $\log Z_{\rm PPISN} - \log Z_{\rm PLP} = {14.3\pm0.1}$.~The most relevant quantity is $\mbhmg$, which we measure to be $\mbhmg={70.09_{-14.72}^{+19.82}}\msun$. However, as is seen in Figure~\ref{fig:spin-corner}, the distribution is bimodal, with the strongest support found at lower values, $\mbhmg \sim 50 \msun$. As noted above, the feature at $\mbhmg$ is a prediction of stellar structure theory, which prevent BHs with masses in the approximate range $50\msun$--$120\msun$ (depending on the stellar parameters) from forming \cite{Farmer:2019jed,Farmer:2020xne,Mehta:2021fgz}.
~In the 1g mass function \eqref{nbh-1g}, the upper mass gap feature is incorporated by virtue of being predicted by stellar structure theory and is therefore detected by design.~It is possible that, when using phenomenological BHMFs that do not separate populations, this feature may be obscured by 2g mass gap \textit{pollutants}.~The posterior we find for $\mbhmg $ is, besides being bimodal, also weakly correlated with $\lambda_{12}$, $\lambda_{21}$. This correlation is expected because for larger $ \mbhmg$, fewer black holes will be found in the higher generation population. 

The posterior for $b$ is well converged 
and not strongly correlated with any of the other variables.~Physically, $b$ is an index corresponding to the slope of the number of stars that collapse to form BHs as a function of mass before accounting for PPISN.~The absence of a peak in the mass function \eqref{nbh-1g} could have implied a less well constrained parameter $b$, if a peak should be there. The fact that $b$ is well constrained indicates that there are a sufficient number of large-mass black holes in the first generation to resolve the power law.~Conversely, the parameter $d$, which physically represents the power law index of the higher generation population, is less well constrained. Future catalogs with more heavy mergers are likely to improve upon this. 
The larger preferred values of $\gamma $ imply a merger‐rate density that grows somewhat more steeply with redshift, suggestive of either shorter delay times or selection effects toward higher-\(z\) systems.

\begin{table}[t]
\centering
\begin{tabular}{|c|c|}
\hline
\multicolumn{2}{|c|}{Mass Function (Eqs.~\eqref{eq:multipop}, \eqref{nbh-1g}, and \eqref{eq:2gMF})}    \\\hline

$\mbhmg               $  &   $ 70.09_{-14.72}^{+19.82}\msun$\\\hline
$a                    $  &   $  0.12_{- 0.08}^{+ 0.13}$\\\hline
$b                    $  &   $ -2.46_{- 0.19}^{+ 0.18}$\\\hline
$\log_{10}\lambda_{21}$  &   $ -2.56_{- 1.57}^{+ 1.34}$\\\hline
$\log_{10}\lambda_{12}$  &   $ -4.77_{- 1.57}^{+ 1.34}$\\\hline
$d                    $  &   $ -6.20_{- 2.03}^{+ 2.01}$\\\hline
$M_{\rm mmin}         $  &   $  3.99_{- 0.97}^{+ 0.96}\msun$\\\hline
$\delta_m             $  &   $  4.42_{- 2.19}^{+ 2.23}\msun$\\\hline
$\beta_q              $  &   $  3.23_{- 0.93}^{+ 1.16}$\\\hline
$\gamma               $  &   $  3.62_{- 1.00}^{+ 1.06}$\\\hline
\end{tabular}%
\caption{Best fit values and error estimates of the mass function analysis.
}
\label{tab:bestfit-massfct}
\end{table}

\subsection{Spectral Siren Analysis}

In the top panel of Figure~\ref{fig:h0corner} we show the marginalized posterior probability distributions for $H_0$ for the two different black hole mass functions introduced above:~the PLP and our astrophysically-informed scenario.
The PLP model yields
$H_0 = 85.37^{+25.18}_{-21.79}\,\mathrm{km\,s^{-1}\,Mpc^{-1}},$
while our analysis gives
$H_0 = 94.62^{+30.24}_{-30.83}\,\mathrm{km\,s^{-1}\,Mpc^{-1}},$
with both posteriors exhibiting broad confidence intervals. 

In Figure~\ref{fig:h0corner} we also show how $H_0$ correlates with $\mbhmg$ in the PPISN model.
~The weakly negative correlation is explained by the fact that, since the luminosity distance and detector frame masses are observed quantities, smaller $H_0$ implies larger source frame masses.
~The values of $H_0$ with $1\sigma$ uncertainties reported by the Planck \cite{Planck:2018vyg} and SH0ES \cite{Riess:2021jrx} collaborations
are consistent with a large range of values of $\mbhmg$, though remarkably not with small values $\lesssim 40 \msun$ at the 2$\sigma$ level. 
We also find that $H_0$ correlates negatively with evolution parameter $\gamma$, consistent with Ref \cite{Pierra:2023deu} who found that the $H_0$ measurement is biased if redshift-dependence is not accounted for.

\section{Discussion}
\label{sec:discussion}

In this study, we examined the population statistics of black holes in the LIGO/Virgo/KAGRA GWTC-3 catalog using a parametric mass function based on simulations of massive stars undergoing pulsational pair-instability supernovae.~The novelty of this approach is that different sub-populations of black holes originating from either stellar collapse or previous mergers can be distinguished, enabling separate measurements of their properties and increasing the amount of information that can be extracted from the data.

Our BHMF provided a better fit to the LVK O3 data than the power law+peak model, and we obtained a measurement of the location of the lower edge of the upper black hole mass gap:~$\mbhmg={70.09_{-14.72}^{+19.82}}\msun$.~This is larger than found in the analysis of GWTC-2 using the same mass function \cite{Baxter:2021swn}. While still in the range of values found through stellar structure simulations, the current preference for a larger value is intriguing because the previous analysis focused exclusively on the primary population, which (by definition) comprises larger masses compared to the secondary population. 
The fact that the power law index of the 1g population $b$ is much better constrained in the present analysis, which may reflect the increased size of the catalog, although the posterior distributions remain relatively broad and, in some cases, bimodal.

In this context, recent analyses of the GWTC-4 catalog have reached differing conclusions regarding the presence and location of the upper mass gap. Some studies report evidence for a gap around $40$–$50\,\msun$ \cite{Tong:2025wpz,Antonini:2025ilj} , while others favor larger values in the range $\sim50$–$90\,\msun$ \cite{Wang:2025nhf}, and still others find no conclusive evidence \cite{Ray:2025xti}. These differences highlight the sensitivity of inferred features to population modeling assumptions. An important avenue for future work is to extend MULTIPASS to GWTC-4, enabling a direct comparison within the framework of our multi-generational, astrophysically-informed mass function.

In this work, we have not explicitly modeled the spin distribution. Instead, we treat the mass function as defined in mass space and marginalize over spin using the posterior samples provided in GWTC-3. 
The spin measurements in GWTC-3 are typically weak \cite{LIGOScientific:2021psn}, and since the features relevant to PISN physics and spectral sirens are primarily encoded in the BHMF, we restrict the analysis to masses for a simpler baseline analysis.
We emphasize, however, that this choice implies that our results should be interpreted as constraints on the mass distribution marginalized over spin. Spin-dependent selection effects are absorbed into this marginalization, and our framework does not attempt to distinguish formation channels on the basis of spin alignment or magnitude. Including spin information will become more important as measurements improve.

Another avenue for future work is incorporating additional astrophysically-predicted features into the BHMF.~For example, reference \cite{Croon:2023kct} predict a peak feature due to $\alpha$-ladder shell burning in the range $35$--$50\msun$, depending on the $^{12}{\rm, C}(\alpha,\gamma)^{16}{\rm O}$ rate). One could then extend our formalism to include this peak using a similar procedure to the derivation of equation \eqref{nbh-1g} whereby one first finds a suitable fitting function for the theoretical peak of \cite{Croon:2023kct} and uses this to determine the corresponding term that must be added to the BHMF.~The detection of this feature would then provide further insights into the processes governing massive stars and black hole formation.

\begin{figure}[t]
    \centering
    \includegraphics[width=.5\textwidth]{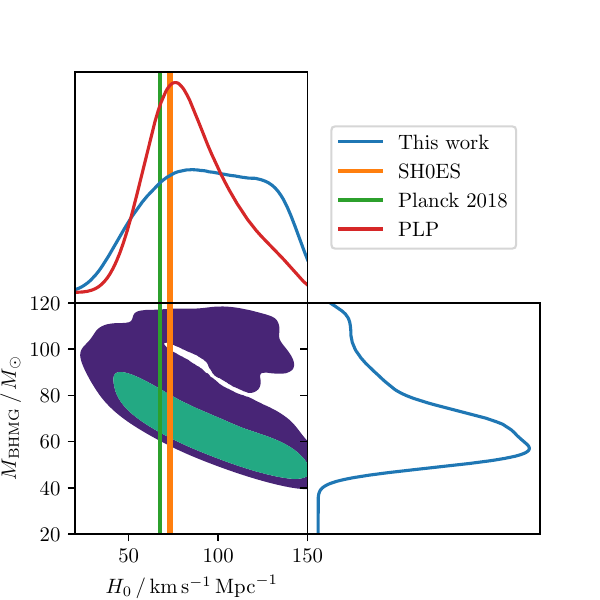}
    \caption{
    Corner plot of $H_0$ vs.~$\mbhmg$ for our astrophysically-informed model \eqref{eq:multipop}. We include the $ H_0$ posterior probability for the PLP model \eqref{eq:plp} and the $1\sigma$ ranges found by the SH0ES and Planck collaborations.}
    \label{fig:h0corner}
\end{figure}

Finally, we applied the spectral siren method for measuring the Hubble constant to our formalism.
We find $H_0 = 94.62^{+30.24}_{-30.83}\,\mathrm{km\,s^{-1}\,Mpc^{-1}},$
in agreement with both the value found by Planck and the value found by the SH0ES measurement.

\section{Conclusions} 

A major goal of gravitational wave astronomy is to interpret the data within the context of fundamental astrophysics.~This requires connecting the black hole mass function with the predictions of stellar physics.~When confronted with current data, our framework for accomplishing this, which enables separate measurements of population properties for different black hole sub-populations, yielded several results compatible with theoretical expectations.
In particular, we obtained a measurement of the location of the upper edge of the black hole mass gap consistent with the predictions of stellar structure theory.

\section*{Software}

Multinest 3.12 ({\tt cc616c1}), Getdist 1.4.7, Astropy 5.3.4, Matplotlib 3.8.0, Numpy 1.26.1, Scipy 1.11.3, h5py 3.9.0

\section*{Acknowledgements}
We thank Eric Baxter and Reed Essick for useful discussions. 
~DC is supported by the STFC under Grant No.~ST/T001011/1.
~This material is based upon work supported by the National Science Foundation under Grant No.~2207880.~Our simulations were run on the University of Hawai\okina i's high-performance supercomputer KOA.~The technical support and advanced computing resources from University of Hawai\okina i Information Technology Services – Cyberinfrastructure, funded in part by the National Science Foundation MRI award \#1920304, are gratefully acknowledged.

\section*{Note added}
Following the submission of this work, the GWTC-3 catalog analyzed here was superseded by GWTC-4 and then GWTC-5, and we have decided not to further pursue journal publication of this analysis. We nevertheless record here, for completeness, several caveats that the reader should bear in mind, some of which were raised during peer review.

Firstly, the overall normalization of the differential merger rate shown in Figure~\ref{fig:masscontours} differs substantially from values reported in the literature, likely reflecting normalization conventions in our pipeline. We have not been able to fully diagnose this discrepancy, and we therefore caution the reader that the relative shapes of the inferred mass functions are more robust than their absolute normalization, and that model comparisons (including the reported Bayes factor) should be interpreted with corresponding care.

Secondly, as discussed in Section V, our marginalization over spin using the GWTC-3 posterior samples implicitly adopts the parameter-estimation sampling prior as the spin population model. Because spins correlate with masses and redshift in individual-event posteriors, this choice may bias the inferred mass distribution. A self-consistent treatment would require including spin explicitly in the population model; our attempts to enlarge the parameter space in this way suffered from convergence problems in the Bayesian inference, potentially relating to computational constraints and the large dimensionality of the parameter space in our model.

These caveats concern the implementation of the inference rather than the physical motivation for the mass function itself, which derives from stellar structure simulations of pulsational pair-instability supernovae and is therefore a credible alternative to purely phenomenological choices.

We thank the anonymous reviewer for raising these issues and leave a full treatment of them, and an extension of {\tt MULTIPASS} to future GWTC catalogues, to future work.

\bibliography{refs}

\end{document}